\documentclass[preprint,aps,prl,showpacs, showkeys,superscriptaddress]{revtex4}
\usepackage{graphicx}
\usepackage{epstopdf}
\usepackage{dcolumn}
\usepackage{bm}
\usepackage{natbib} 
\begin{document}
\title{Dynamic response  of a spin-1/2 Kondo singlet}
\author{Bryan Hemingway} 
\affiliation{Department of Physics, University of Cincinnati, Cincinnati, OH 45221, USA}
\affiliation{Department of Physics and Center for Nanophysics and Advanced Materials, University of Maryland, College Park, Maryland 20742, USA}
\author{Stephen Herbert}\affiliation{Physics Department, Xavier University, Cincinnati, Ohio 45207, USA}
\author{Michael Melloch}\affiliation{School of Electrical and Computer Engineering, Purdue University, West Lafayette, Indiana 47907, USA}
\author{Andrei Kogan}
\email{andrei.kogan@uc.edu}
\affiliation{Department of Physics, University of Cincinnati, Cincinnati, OH 45221, USA}
\date{\today}

\begin{abstract}
We present a study of spin 1/2 Kondo singlets in single electron transistors under a microwave frequency bias excitation. We  compare time-averaged  conductance $G$ to predicted universal response with respect to microwave frequency, oscillation amplitude and the Kondo temperature and find a non-adiabatic response when the  microwave photon energy  $hf$  is comparable to the Kondo temperature $k_B T_K$. We show that our measurements are qualitatively consistent with the predictions for the radiation-induced decoherence rate of the Kondo spin. 
\end{abstract}
\maketitle

The discovery of the Kondo effect \cite{kondo:64,hewson:93,coleman:07} in a Single-electron transistor \cite{goldhaber:n98,goldhaber:98PRL,cronenwett:98} has enabled  precision studies of many-body effects away from equilibrium. The Kondo singlet, formed by a spin confined within a conductor and interacting with itinerant electrons, is a fundamental example of a many-body correlated state which produces a sharp enhancement of the density of states near the Fermi energy and is characterized by the "Kondo Temperature" $T_K$, the effective  many-body spin singlet binding energy \cite{hewson:93}. In a single-electron transistor (SET), an electronic puddle  is confined to a small region (a quantum dot) and contacted via tunnel barriers by two large conducting regions, the drain and the source. The Kondo effect in such a device presents itself as a sharp peak  in differential conductance  $dI/dV_{ds}$  as function of the drain-source bias $V_{ds}$ across the dot. $dI/dV_{ds}$ reaches its maximum value at $V_{ds}=0$, a direct manifestation of the Kondo-enhanced density of states in the dot-leads system. 

Characteristic of the Kondo state is the universal dependence of macroscopic observables on $T/T_K$ at low temperatures \cite{hewson:93}. SETs enabled precision universality tests \cite{goldhaber:n98,goldhaber:98PRL,cronenwett:98}.  Studies of the static conductance of SETs under source-drain  bias have shown that universality is preserved when the deviation from equilibrium is small \cite{grobis:08,liu:09, kretinin:11}, and departures from universality have been observed in SETs subjected to an external magnetic field above a certain threshold value of the Zeeman energy \cite{liu:09,wright:11}. Our  present work is motivated by the efforts to understand the {\em dynamics} of Kondo correlated states, which can be probed with SETs by  varying the spin confinement potential, or bias, or both, in time 
\cite{elzerman:00,kogan:science,delbecq:11}. 

In the time domain, $h/k_B T_K$, where $h$ is the Planck's constant,  is predicted to be a measure of the time that it takes for the Kondo state to develop following a rapid shift of the device parameters into the Kondo regime \cite{nordlander:00}. In the frequency domain, this implies that   that the response of a Kondo singlet to oscillating fields should become non-adiabatic and sensitive to the frequency, $f$, when $f$ is comparable to the "Kondo frequency" $f_K \equiv k_B T_K/h$  \cite{schiller:96,ng:96,lopez:98,goldin:98}. A detailed analysis of transport across an SET when microwave bias or gate voltage signal is present \cite{kaminski:99,kaminski:00} predicts non-adiabatic features in the SET differential conductance which are universal with respect to the drive frequency and the Kondo temperature and reflect a competition between Kondo correlations and radiation-induced spin coherence breaking. Suppression of the Kondo effect  with a gate-coupled microwave signal has been reported \cite{elzerman:00}, and non-adiabatic response evidenced by additional peaks in the nonlinear conductance has been observed \cite{kogan:science}. A recent experiment by Latta et al \cite{latta:11} presents evidence of universality with respect to the ratio of the photon frequency and the Kondo temperature in the absorption spectrum in Kondo-correlated quantum dots. Overall, however, both the phenomenology and theory of a driven Kondo system remain far from complete.

The SETs used in this work are defined via e-beam and photolithography  on a GaAs/AlGaAs heterostructure (Fig.1). The electron density of the 2DEG is $n_{2D} = 4.8\times 10^{11}$ cm$^{-2}$ and the electron mobility is $\mu\ge 5\times 10^5$ cm$^2/$V\,sec, as determined by magneto-transport measurements at $\sim$ 4.2K\cite{liu:09, liu:12}.  The differential conductance is recorded via a lock-in detection of the output of a current preamplifier (DL Instruments 1211)  in series with the SET drain-source path with a modulation signal of 17 Hz and 1.9 $\mu V_{rms}$ added to the SET bias.  The quantum dot is formed by applying negative voltages $V_T$, $V_S$, $V_B$, $V_G$ to the "Top", "Side", "Bottom" and "Gate" surface gates, and the bound electron energy is varied by tuning   $V_G$.  The SET is placed in the  Kondo regime and the microwave-frequency bias oscillations are added to the static SET bias $V_{ds}$ so the total bias voltage becomes  $V(t)= V_{ds}+V_{AC} \cos 2\pi f t$ (Fig. 1).  All data obtained with microwaves are recorded with $V_{ds}=0$.   The SET conductance $G$ is defined as the derivative of the time-averaged source-drain current with respect to the static portion of the bias $V_{ds}$
\begin{equation}
G=d\langle I\rangle/dV_{ds}
\end{equation}

where $\langle \dots\ \rangle$ indicates  averaging over a large number of bias oscillation cycles.  The averaging time is determined by the time constant of the DC current measurement circuit, and is of order of 0.3-1 seconds. A summary of symbols used to refer to the SET conductance in different regimes and other symbols is given in table \ref{tab:1}.

Our goal is to identify non-adiabatic, frequency-dependent features in $G$ expected to emerge at frequencies $f$ comparable to $f_K$. First, we tune the SET to an odd occupancy, noted by the emergence of the strong zero bias Kondo conductance peak in a Coulomb-blockade valley flanked by Kondo-free valleys (Fig. 2 (a), (b)).  To obtain the Kondo temperature, we   measure the static zero-bias conductance $G_0=G_{DC}(V_{ds}=0)$ as function of $V_G$ across the Kondo valley in a wide temperature range, and then fit the data to an empirical scaling function \cite{goldhaber:n98} known to represent equilibrium RG calculations. Details of this procedure are described in the Supplemental Material \cite{Microwave_Supplement}. With the Kondo temperatures at the center of the valley determined (0.42 K in SET1 and 0.67 K in SET2), we fix $V_G$ so that the static gate voltages keep the device at the same Kondo temperature for all microwave measurements. Possible oscillations of $T_K$ due to a leakage of microwave signal from the source and drain onto the gates is minimized because the Kondo temperature is at its minimum with respect to the dot level energy.  The  microwave-frequency oscillating bias voltage is supplied via a multi-section coaxial line  equipped with heat sinking stages for the inner conductor at the 1K pot and the mixing chamber and coupled capacitively to the device leads. The base electron temperature in our dilution refrigerator is approximately 55 mK \cite{liu:09} and rises to  70 mK  with the coaxial line coupled to the sample (determined  from the sharpness  of co-tunneling features  across a Zeeman-split level in the  weak-coupling limit \cite{liu:12}), significantly below the Kondo temperature in both devices used for the experiment.   The samples show excellent long term stability, with the  zero bias static Kondo  peak varying  by less than 5\% over several months of measurements, based on frequent checks.

Figure 2 (a), (b) shows maps of static nonlinear conductance $G_{DC}$ for the Kondo states in SET1 and SET2. The dashed lines indicate values of the gate voltages $V_G$ selected for the measurements. The corresponding Kondo peaks are shown in Figure 2 (c), (d). The Kondo enhanced conductace at zero bias is superposed over a broad conductance minimum at $V_{ds}=0$. As $V_{ds}$ is tuned away from zero, the conductance first drops and then  rises slowly as the energy of the Coulomb-blockaded level in the dot is pulled closer to the window of energies between the drain and the source lead due to the dot-lead capacitance. This gives rise to the well-known "diamond" like features in $G$ plotted as function of $V_{ds}$ and $V_G$. 

We define  the adiabatic regime as the condition when the current across the SET tracks with the time-varying bias according to the static current-voltage response, which is naturally expected to occur at low frequencies. Using the measured static ($V_{AC}=0$)  $G_{DC}(V_{ds})$, one can  calculate $G(V_{AC})$ as follows:

\begin{equation}
G(V_{AC})= \frac{1}{T}\int^T_0 G_{DC}(V_{ds}+V_{AC} \sin (\omega t))dt,
\end{equation}
where $\omega$=2$\pi$f and T is the period of the oscillation. Eq. 2 suggests that the dependence $G(V_{AC})$ in the adiabatic limit does not depend on frequency. Fig 2 (e), (f) schematically illustrates features in $G$ on $V_{AC}$ (with $V_{ds}=0$ maintained) in the adiabatic limit.  When the static source-drain bias $V_{ds}=0$ and an oscillating bias $V_{AC} \cos \omega t$ is present, the time-averaged $G$ is expected to  first decrease as the bias oscillations begin to sample the sides of the Kondo peak (regime 1), then reach a minimum (regime 2) and  to rise again as $V_{AC}$ becomes larger than the Kondo peak width and begins to reach the regions where the static conductance rises (regime 3).

A detailed theory for the non-adiabatic SET response was developed by Kaminski et. al. \cite{kaminski:00} who identify a set of conditions  for the onset of dynamics which depend on  the relationship between $f$, $V_{AC}$ and $T_K$ and derive expressions for $G$ at frequencies $f > f_K$ .  Figure \ref{kng}, reproduced from \cite{kaminski:00},  presents a "map" of the relevant parameter space, in which the microwave photon energy and the bias oscillation amplitude are scaled by the Kondo temperature. The numbers in the map refer to equations for $G$ presented in \cite{kaminski:00}. For convenience, we give a summary of equations (72 -77) in \cite{kaminski:00}  and discuss their meaning, and refer the reader to the original work for further discussion and derivations.

At low frequencies, $f<f_K$, as well as at high frequencies but with sufficiently large excitation bias amplitude, Equations (76) and (77) from \cite{kaminski:00} the peak conductance $G_0$ is given by:

\begin{equation}
Eq. (76)~~~ G_0=\left\{1-\frac{3}{16}\left(\frac{eV_{AC}}{T_K}\right)^2\right\} G_U
\end{equation}

\begin{equation}
Eq. (77)~~~ G_0=\frac{3 \pi^2}{16}\frac{1}{[\ln (eV_{AC}/T_K]^2]} G_U
\end{equation}

where the  coefficient $G_U$ is the static conductance, which depends on the asymmetry between the QD and the two electron reservoirs and can be measured directly. Here and below we have changed several symbols to maintain consistency with the notations used in this paper (table \ref{tab:1}). It is important that both are derived by time-averaging of the static response

\begin{equation}
Eq. (75)~~~ G_{0}= \overline{G(V_{AC} \cos \omega t})
\end{equation}
which is mathematically equivalent to our Eq. 2, and thus describe the adiabatic limit.  For small amplitudes and in the dynamic regime, $f>f_K$, Eq. (74) 

\begin{equation}
Eq. (74) ~~~ G_{peak} = (1-a\frac{\hbar}{\tau k_B T_K})G_U
\end{equation}

predicts the conductance to be linear with respect to the effective spin decoherence rate:

\begin{equation}
Eq. (72) ~~~\frac{\hbar}{\tau k_B T_K} = \frac{1}{\pi}\frac{G_U}{e^2/\hbar}(\frac{eV_{AC}}{k_BT_K})^2\frac{k_BT_K}{hf}\frac{1}{ln(hf/k_BT_K)^2} \label{eq:tau}
\end{equation}

where the coefficient $a$ is estimated theoretically to be of order 1. Last, Eq. (73) (see also Fig. 4 in \cite{kaminski:00}) describes the intermediate regime between the non-adiabatic (74) and the adiabtic (eq. 77) limits as the high frequency bias amplitude $V_{AC}$  grows. 
\begin{equation}
Eq. (73)~~~ G_0= \frac{3 \pi^2}{16} \frac{1}{[\ln (h/\tau T_K)]^2} G_U
\end{equation}

To summarize, it is predicted that 
\begin{enumerate} 
\item transport across  a Kondo-correlated SET should be adiabatic not only for $f\ll f_K$ but also for $f \ge f_K$ as long as the AC bias amplitude is large on the scale of $k_B T_K/e$. The latter can be understood  as the gradual disappearance of the slow Kondo dynamics far from equilibrium, at large voltages predominantly sampled by the large-amplitude AC bias.
\item at small excitation amplitudes, the  conductance should be rising with frequency  according to Eq. 6 ( Eq (74) in \cite{kaminski:00}) when the dynamic regime is reached. This effect stems from the predicted reduction in the radiation-induced decoherence rate with frequency according to Eq. \ref{eq:tau} and has not been observed previously. 
\end{enumerate}

The aim of this paper is to test these predictions. We select several frequencies spanning the range from 1 GHz to 30 GHz to cover both $f<f_K$ and $f>f_K$ limits. For each frequency, we record $G$ at $V_{ds}=0$ as function of microwave power ( Figure 4) and compare the shape of the $G(V_{AC})$ data to the expected response in the adiabatic regime. The adiabatic $G(V_{AC})$ response for each state is obtained from measured static nonlinear conductance $G_{DC}$ (Fig. 2 (c), (d)) via  Eq. 2.

At each microwave frequency, measurements of $G$ versus power  reveal a drop in $G$ followed by an upturn that occurs at  large powers (Figure 4). Qualitatively, this behavior resembles the expected adiabatic response (Fig 1 (e), (f)). A closer inspection of the data, however shows that the rate of the drop of $G$ with power at moderate powers is sensitive to frequency, which is indicative of the onset of non-adiabaticity. To quantify this sensitivity to frequency of the observed response and perform comparisons to theory, one first needs to understand factors that define the relationship between the microwave power at the sample contacts and the microwave-frequency voltage produced across the SET island. In our experiment, as often is the case in microwave measurments with SETs, this relationship cannot be determined by an independent measurement. It makes it impossible for us to test the two theoretical predictions 1 and 2 independently. Instead, we aim for a more limited goal of testing their mutual consistency.

First, we note that the  current flow across the SET has no  effect on the microwave field distribution in the coaxial line, the sample-to-coax coupling and the sample because the effective SET impedance, of order 10-100 k$\Omega$, is at least two orders of magnitude higher than the geometric impedances of the line and the line-sample coupling.  Therefore, the resistive component of the SET impedance presents an "infinite load" to the microwave circuit. Since the drive voltages used in the experiments are very small, less than 1 mV, we are justified to ignore nonlinearities in the microwave delivery path with respect to the microwave power. The relationship between the microwave power and the microwave SET bias modulation amplitude $V_{AC}$ can therefore be described by a linear form 
\begin{equation}
V_{AC}=T(f) P^{1/2} 
\end{equation}

where the coefficient $T(f)$ is an unknown function of frequency only. Elzerman et. al. \cite{elzerman:00} proposed a method for $V_{AC}$ determination based on a comparison of the effective microwave voltages at the device to the electron temperature, but known ambiguities with the interpretation of results introduced by this approach \cite{elzerman:00, kaminski:00} limit its usefullness. Here, we first assume that the predicted recovery of the adiabatic regime at high large AC powers, is valid. This implies that the microwave voltage which is large enough to produce the upturn in $G(V_{AC})$ is also large enough to produce an adiabatic response, consistent with the predictions \cite{kaminski:00}. This assumption is justified further  by examining the slopes of $G(V_{AC})$ curves in the upturn sections, which are remarkably similar to the one found in the adiabatic calculation. We therefore replot the $G$ power dependence data by choosing a $T(f)$ value at each frequency so that the measured conductances at different frequencies agree numerically at the larger amplitudes with the adiabatic calculation, as shown in Figures 5 (a) and 6 (a). Compared to the adiabatic case, as the frequency increases, $G(V_{AC})$ begins to fall at lower $V_{AC}$, and the drop in $G(V_{AC})$ is noticeably more gradual, with measured $G$ values  falling systematically below the adiabatic curve (solid black line). Interestingly, this difference becomes most prominent as the frequency approaches values of order $k_B T_K/h$. This is illustrated in Figures 5 (b) and 6 (b) where we plot $G$ as function of frequency at a fixed $V_{AC}= 20 \mu V$ where the deviation from the adiabatic curve is strong in both SETs. As the frequency increases above $f_K$,the conductance rises again, which results in  a "dip" in $G$ at $f \sim f_K$ for both Kondo states. Thus, our measurements are qualitatively consistent with both the predicted recovery of the adiabatic response at large $V_{AC}$ and the increase in $G$ with frequency in the dynamic regime, $f>f_K$.  

The rise in dot temperature due to Joule heating is small and does not influence $G$ measurements significantly. Thermal conductivity of the 2DEG can be estimated from the 2DEG parameters via the Wiedemann-Franz Law and is approximately $10^{-10}$ Watts/Kelvin.  Neglecting electron-phonon interactions and restricting the power flow to the electrons in the device mesa only (worst-case) one finds that bias oscillation with amplitude $V_{AC}= 20 \mu$V  will raise the dot temperature by 0.3 mK, which increases to $\sim$7 mK for $V_{AC}= 100 \mu$V. Significantly larger temperature increases would be needed to produce conductance suppression comparable to that shown in Figures 5 and 6. 

A quantitative comparison to theory is presented in Figure 7 where we plot $G$ scaled by $G_0 = G(V_{AC})=0$ for low and moderate $V_{AC}$ where the dynamic response is expected to dominate. The dotted lines show a calculation based on Eqs. 6 and 7. An approximate scaling at sufficiently high frequencies is apparent in the data from SET2 ( Figure 5 (b) ), but the scaling fails in SET1. In both devices, the decrease in $G$ with increasing spin decoherence rate $1/\tau$ is nearly logarithmic and is noticeably slower than the theory suggest. Again, it is important to emphasize that the numeric comparison of $G$ at a given value of $V_{AC}$ and $1/\tau$ is based on the assumption that the upturn of $G$ at high powers is governed by adiabatic, frequency-independent physics not sensitive to intrinsic Kondo dynamics and its intrinsic frequency scales $f_K$. A further test of the theory requires an independent measurement of $V_{AC}$ as function of microwave power. This can be achieved by using the SET as a microwave detector in a regime with known dynamics, such as photon-assisted tunneling across the dot weakly coupled to the leads at moderate temperatures so that the Kondo physics is suppressed. Alternatively, the sample could be fitted with an independent microwave detector in parallel with the SET. These measurements will be presented elsewhere.

In conclusion, we have presented measurements of the time-averaged SET conductance $G$ in the Kondo regime with  a high-frequency excitation $V_{AC} \cos(2\pi f t) $  added to the  zero static SET bias $V_{ds}$. In the two SET samples investigated, we found frequency-dependent $G(V_{AC}$ and showed that the deviation of $G(V_{AC})$ from the adiabatic response is most prominent at frequencies comparable to $f_K$.  We have also  shown  that the predicted decrease in the spin decoherence rate at high frequencies $f\ge f_k$ is consistent with the recovery of the adiabatic regime at  excitation amplitudes exceeding $k_B T_K/e$ \cite{kaminski:00}. In one of the two devices investigated, we found limited scaling of $G$ the spin decoherence rate and the Kondo temperature. We also report a systematic discrepancy between the theory and the measurements. We hope that these observations will further stimulate both theoretical studies of the dynamic Kondo effect, especially  at moderate frequencies $f\le f_K$ where currently there are no specific predictions, and experimental studies with a better control of the microwave AC bias voltage magnitude.

The research is supported by the NSF DMR award No. 0804199, NSF DMR award No. 1206784 and by University of Cincinnati.  We thank L. Glazman for fruitful discussions, Tai-Min Liu for his help in SET fabrication, and J. Markus, M. Ankenbauer and R. Schrott for technical assistance.

\bibliographystyle{apsrev}

\begin{table}
\renewcommand{\arraystretch}{1.3}
\begin{tabular}{|c c p{15 cm} |}
\hline
\hline
 & &\\
 $V_G$ & & DC voltage on gate\\
 $V_{ds}$ & & DC drain-source bias\\
 $V_{AC}$ & & Bias oscillation amplitude\\
 \hline
 $G_{DC}(V_{ds})$ & & The static nonlinear conductance $dI/dV_{ds}$ with  $V_{AC}=0$.   \\
 $G(V_{AC})$ && The derivative of time-averaged current $d \langle I\rangle/d V_{ds}$ at $V_{ds}=0$ with microwaves present ($V_{AC}\ge 0$)\\
 $G_0$ & & The peak value of $G_{DC}$ at $V_{ds}=0$ and base electron temperature. Same as $G$ at $V_{AC}=0$. In the supplement, $G_0(T)$ refers to the zero-bias conductance as function of temperature $T$  \\
 $G_U$ & & Static SET conductance in the $T=0$ limit \\
\hline 
$T_K$ &&  Kondo temperature \\
 $ f$ & & bias oscillation frequency  \\
 $\omega$, $\omega^\prime$ & &$2\pi f$, angular bias oscillation frequency. $\omega^\prime$ is the notation used in \cite{kaminski:00} \\
 $f_K$ & & The Kondo frequency scale $k_B T_K/h$, where $k_B$ is the Boltzmann constant and $h$ is the Planck's constant.  \\
\hline 
$\tau$ & &Inverse rate of spin decoherence caused by microwave \\
 $\Gamma$ && dot-lead tunneling rate (see Supplement).  \\
 $U$ && dot charging energy. \\
 $\epsilon$ && dot level energy, measured from the charge degeneracy point at the boundary of the Kondo valley. \\
\hline
\end{tabular}
\caption[table]{\label{tab:1}Symbols and notations}
\end{table}

\newpage

\begin{figure}
\includegraphics[width=3in, keepaspectratio=true]{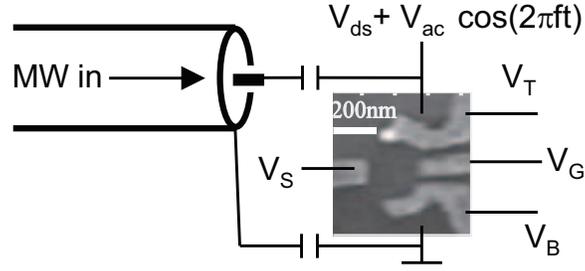}
\caption[circuit]{\label{fig:1} Electron micrograph of the SET  showing the labeling of DC voltages on the ``Side", ``Top", ``Bottom", and ``Gate" gates that define the SET.  The oscillating voltage with amplitude $V_{AC}$ and microwave frequency $f$ is supplied by a coaxial line  coupled capacitively to the source-drain path. The DC (static) portion of the SET drain-source bias, $V_{ds}$, is supplied by a separate filtered computer-controlled DC circuit (not shown).}
\end{figure}

\begin{figure}
\includegraphics[width=4in, keepaspectratio=true]{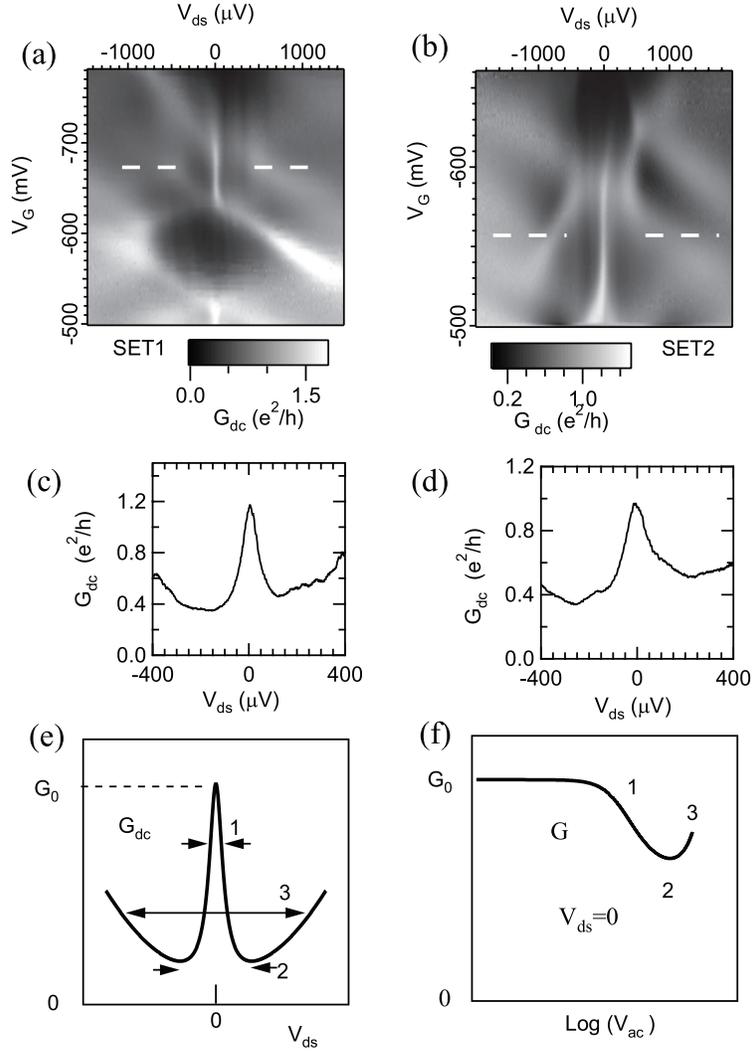}
\caption{\label{fig:2} (a), (b): Differential conductance map in SET1 (a) and SET2 (b). Dashed lines indicate the gate voltages $V_G$, -673 mV for SET1 and -558 mV for SET2, used in the experiment. (c), (d): corresponding static conductance $G_{DC}$ traces. (e), (f) a Schematic illustrating three different regimes depending on the excitation voltage $V_{AC}$ in the adiabatic limit, and the corresponding changes in the zero-bias time-averaged conductance $G$ (1) Low $V_{AC}$: $G$ drops with increasing $V_{AC}$, reaches a minimum (2) and begins to increase again (3). For explanations, see text.}
\end{figure}

\begin{figure}
\includegraphics[width=3.4in, keepaspectratio=true]{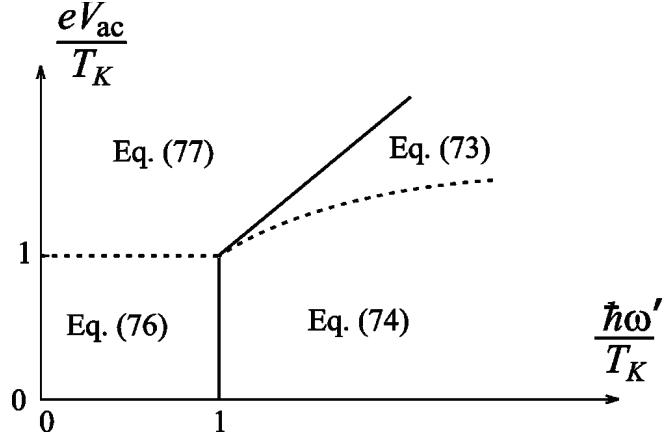}
\caption{\label{kng} Predicted boundaries between the adiabatic and non-adiabatic regimes in a Kondo-correlated SET at zero static bias and different AC bias amplitudes $V_{ac}$ and bias oscillation frequencies $\omega^\prime$.  Reproduced with permission from ref. \cite{kaminski:00}, Figure 3. Eq. (76) and (77) in \cite{kaminski:00} describe the adiabatic regime and are derived from a form mathematically identical to  Eq. 1 in this paper.  Eq. (74)  (Eq. 2 in this paper) and Eq. (73)  in \cite{kaminski:00} describe the  high frequency non-adiabatic regime.}
\end{figure}

\begin{figure}
\includegraphics[width=5in, keepaspectratio=true]{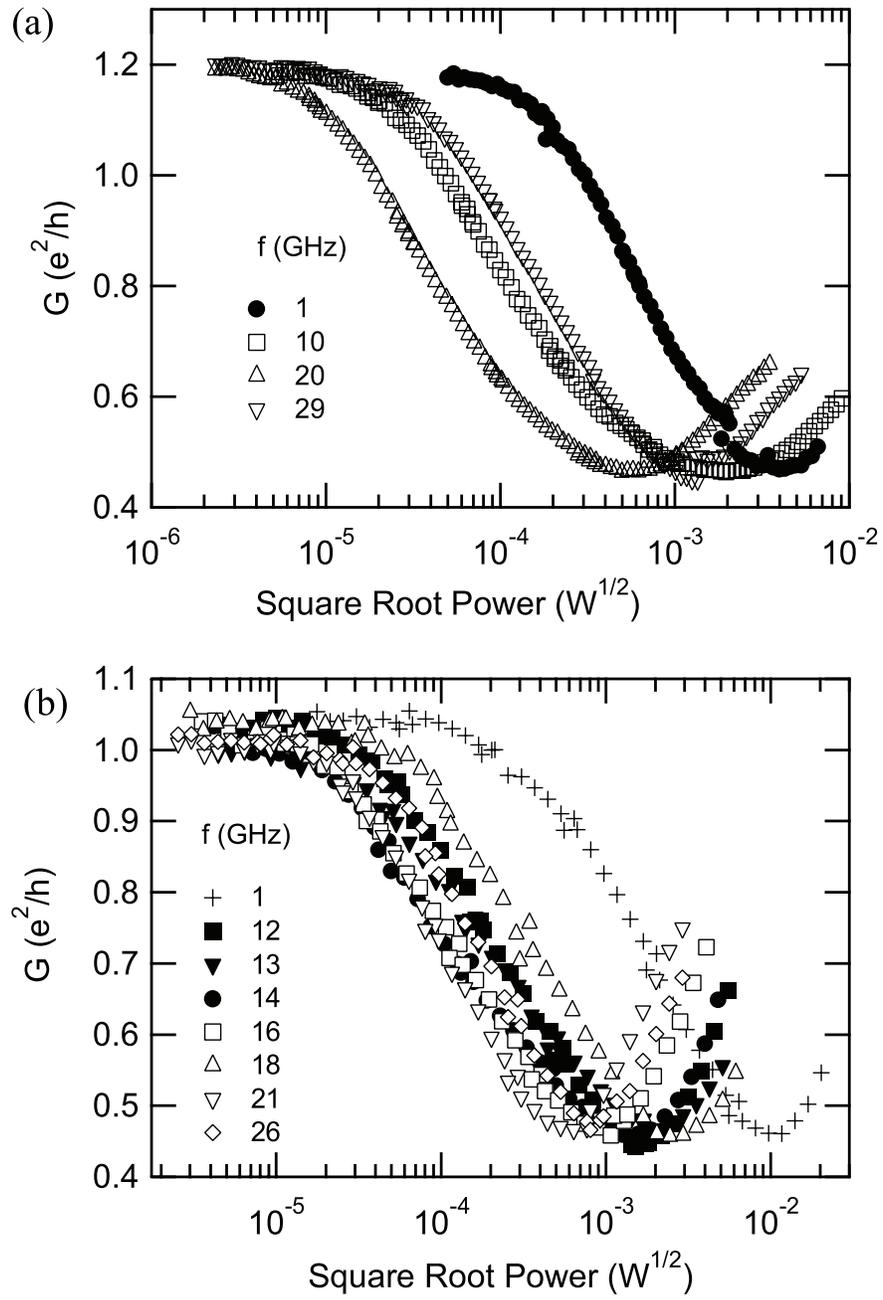}
\caption{\label{fig:4} Zero-bias conductance $G$ as function of power at the end of the microwave line. (a) SET 1. (b): SET2}
\end{figure}

\begin{figure}
\includegraphics[width=4.5in, keepaspectratio=true]{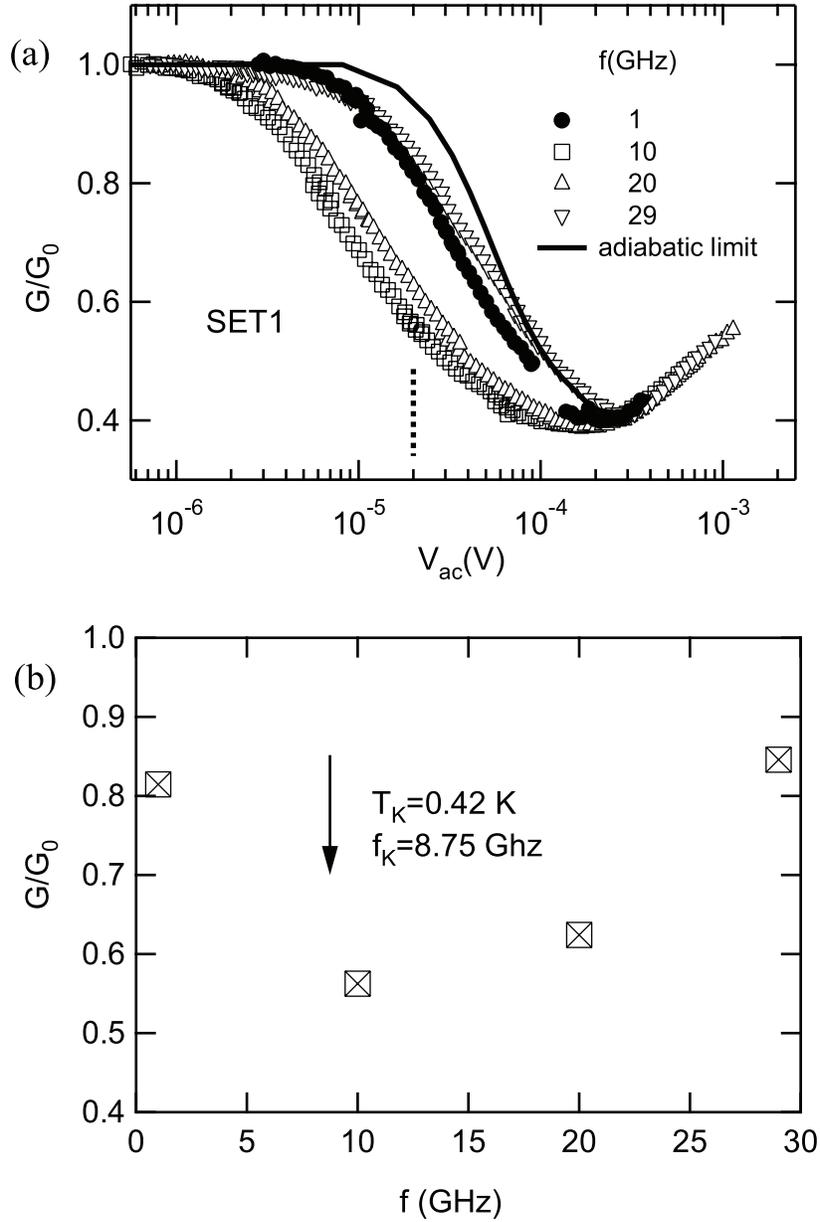}
\caption{\label{fig:5} (a) Conductance as function of $V_{AC}$. Conversion from microwave power to $V_{AC}$ for each frequency is done by matching the high-power rise of $G$ , regimes (2) and (3) shown in Fig 2 (e), (f), to the adiabatic calculation (solid black line). (b): Conductance as function of frequency at a fixed  $V_{AC}= 20 \mu V$. The $V_{AC}$ value ( (a), dashed line) is  chosen in the range where the curves show the strongest sensitivity to microwave frequency. Vertical arrow marks frequency $f_K$ at which the photon energy matches the Kondo temperature, $f_K = k_B T_K/h$. The conductance suppression is maximized at frequencies of order $f_K$. For the analysis of Kondo temperatures, see Supplemental Material. }
\end{figure}

\begin{figure}
\includegraphics[width=4.5in, keepaspectratio=true]{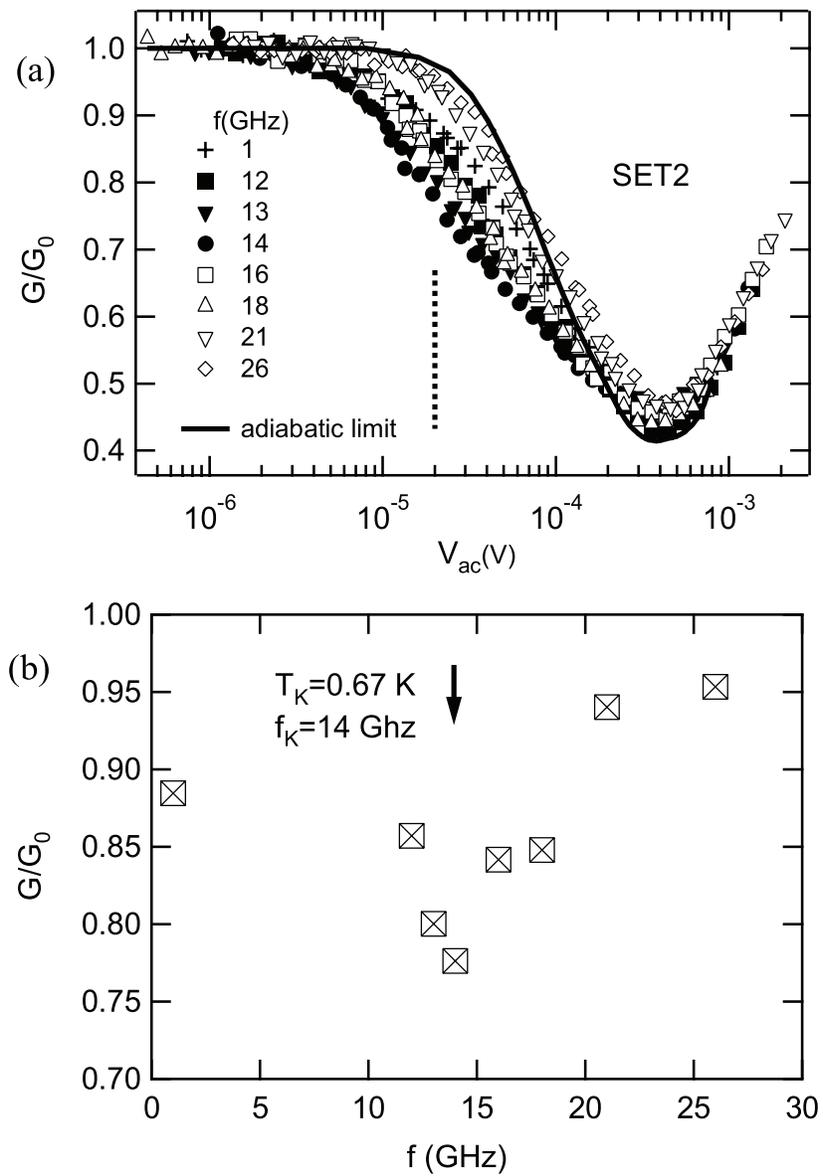}
\caption{\label{fig:6}  (a) Conductance as function of $V_{AC}$ for SET2. Figure notations are same as for SET1 (See caption for Figure 5)}
\end{figure}

\begin{figure}
\includegraphics[width=4.5in, keepaspectratio=true]{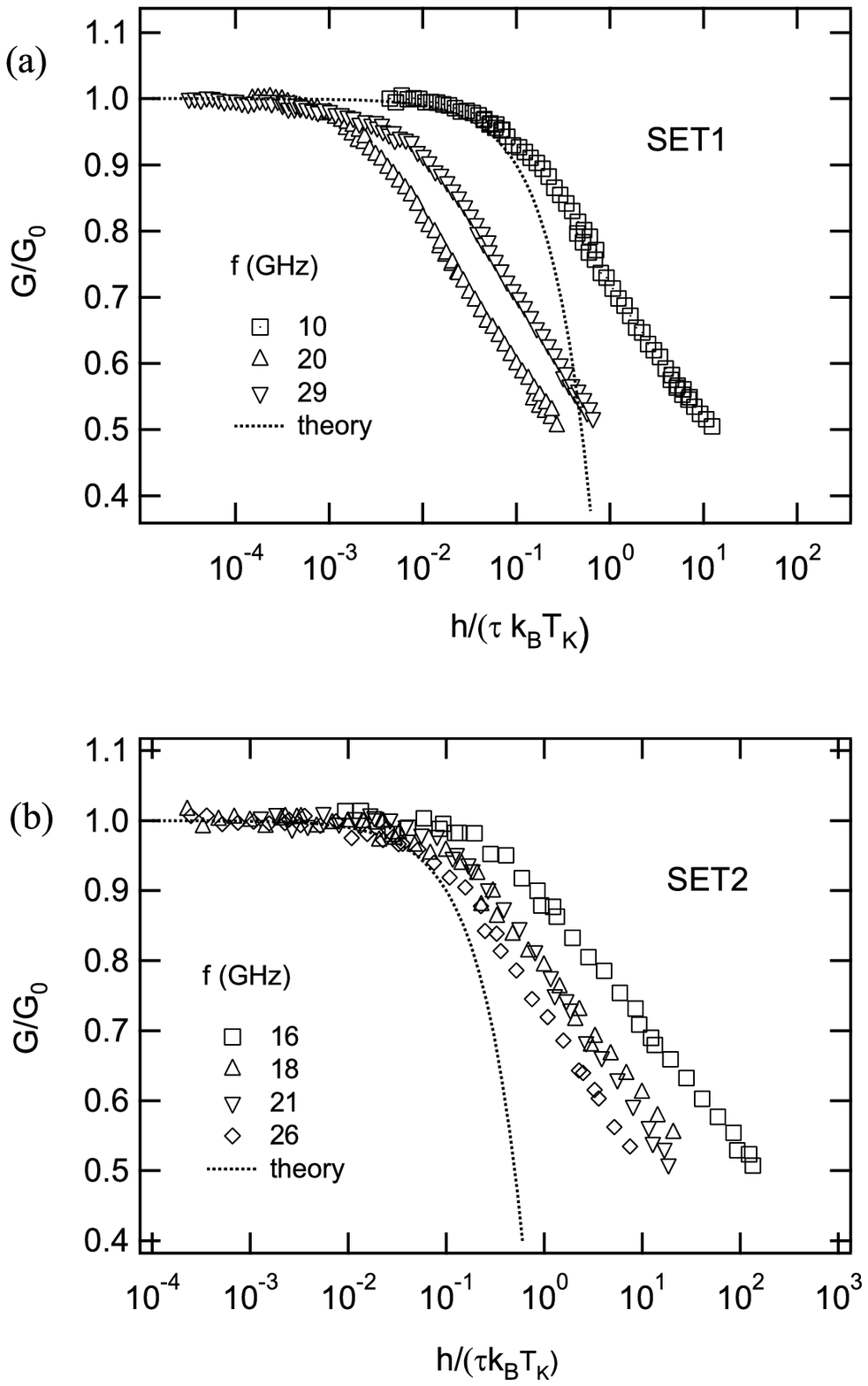}
\caption{\label{fig:7} Universality check: peak conductance $G(V_{AC})$ scaled by $G_0$ as function of the predicted spin decoherence rate, scaled by the Kondo temperature. for SET1 (a) and SET2 (b). The dotted lines show the predicted universal dependence of $G$ on the spin decoherence rate $\tau$ .  }
\end{figure}

\end{document}


\title{Dynamic response  of a spin-1/2 Kondo singlet: Kondo temperature measurements. Supplemental material}
\author{Bryan Hemingway} 
\affiliation{Department of Physics, University of Cincinnati, Cincinnati, OH 45221, USA}
\affiliation{Department of Physics and Center for Nanophysics and Advanced Materials, University of Maryland, College Park, Maryland 20742, USA}
\author{Stephen Herbert}\affiliation{Physics Department, Xavier University, Cincinnati, Ohio 45207, USA}
\author{Michael Melloch}\affiliation{School of Electrical and Computer Engineering, Purdue University, West Lafayette, Indiana 47907, USA}
\author{Andrei Kogan}
\email{andrei.kogan@uc.edu}
\affiliation{Department of Physics, University of Cincinnati, Cincinnati, OH 45221, USA}
\date{\today}
             \maketitle
             
To determine the Kondo temperature in a given Kondo state, we measure the conductance $G_0(T)$ at zero $V_{ds}$ bias as function of gate voltage $V_G$ and temperature $T$ across the Kondo valley.  In this Supplement, we define $G_0(T)$ as the no-microwaves, temperature-dependent Kondo peak height \footnote{To avoid possible confusion, we point out that, in the main text, the zero-bias conductance $G_0$ is always taken at the base electron temperature (See Table 1 in the main manuscript).}. The procedure for the Kondo temperature determination is the same as in our previous work \cite{liu:09}. For a spin 1/2 Kondo state, different $V_G$ give different Kondo temperatures $T_K$ according to the Haldane-like formula:
\begin{equation}
T_K = A \frac{\sqrt{\Gamma U}}{2} \exp[\pi \epsilon(\epsilon+U)]/\Gamma U
\label{eq:haldane}
\end{equation}
Where $U$ is the dot charging energy, $\Gamma$ is the effective dot-lead tunneling rate and $\epsilon$ is the dot energy measured from the charge degeneracy point. We have added an empirical prefactor $A$ typically found to be of order one with both $A<1$ and $A>1$ reported in literature \cite{Goldhaber:98PRL, liu:09}.  $\epsilon$ is related to the gate voltage by
\begin{equation}
\partial \epsilon /\partial V_G=-\alpha_G
\end{equation}
where $\alpha_G$ is the ratio of the capacitance of  the voltage-swept gate to the total capacitance between the dot and the surrounding gates and dot contacts. The values of $\alpha_G$ are determined from the slopes of the Coulomb Blockade diamonds. $G_0 (T)$ data scaled by the zero-temperature conductance limit $G_U$ \footnote{Note that $G_U$ depends on the device symmetry and, consequently, the gate voltage $V_G$ and  is generally less than the so-called {\em unitary limit} value $2e^2/h$, only achievable if the source and drain barriers have identical transmission coefficients.},  plotted  as functions of the ratio $T/T_K$, should collapse onto a universal curve given, for a spin 1/2 Kondo state, by

\begin{equation}
G_0(T)/G_U = \left(\frac{T_K^{'2}}{T^2 + T_K^{'2}}\right)^s
\end{equation}
\begin{equation}
T_K^{'2} = \frac{T_K^2}{2^{1/s}-1}
\end{equation}

so that the Kondo temperature has the meaning of temperature $T$ at which the conductance is  $G_U/2$. The numeric parameter $s$ determines the shape of the curve and depends on the dot spin; for spin 1/2 the appropriate choice is  $s=0.2$ \cite{Goldhaber:98PRL}. 

A summary of the parameters of the two Kondo states presented in the paper is given in Table \ref{tab:1}. In the absence of the Kondo effect, the tunneling rate $\Gamma = \Gamma_S + \Gamma_D$, where $\Gamma_S$($\Gamma_D$) are the tunneling rate between the dot and the source(drain) contact, determines the width of the Coulomb Blockade peak at the single-particle charge degeneracy value of $V_G$. In a Kondo state, we estimate $\Gamma$from the conductance maps ( Figure 2 (a), (b) in the main manuscript)) at large $V_{ds}$ where the Kondo enhancement does not play a significant role yet the separation of the diamond "edges" does not produce a significant broadening of the conductance peak. From the $\Gamma$ values determined this way, one can estimate the ratio of the Kondo temperature obtained from the fits to the expression in the prefactor of the Haldane formula. Values of 0.32 and 0.88 found in the present work (Table \ref{tab:1}) are in good agreement with prior measurements in SETs. 

Figures 1 and 2 present results of such analysis for the two Kondo states in SET1 and SET2 that were investigated in the paper. Figure 1, a) shows the conductance $G_0(T)$ for several representative gate voltage choices across the valley. The Kondo temperature $T_K(V_G)$ for each $V_G$ and the corresponding $G_U$ are treated as adjustable parameters determined by fitting the $G_0(T)$ to eqs. 3 and 4. Having determined  $T_K(V_G)$ and $G_U(V_G)$, we replot the data in the scaled form  (Fig. 1b) and 2b)) and find excellent scaling in agreement with the eqs. 3 and 4 (dashed lines). Figure 1c) shows $G_0$ across the valley  measured at the base temperature of the dilution refrigerator (left axis) , and the Kondo temperature obtained from the analysis (right axis). We further check the results of the analysis by comparing the $T_K$ values across the valley to the predictions of the  Haldane formula (solid black line)  and find  good agreement in the center of the valley, where the microwave measurements take place.

\begin{table}
\renewcommand{\arraystretch}{1.3}
\begin{tabular}{|c|  c c  c  |c   c|}
\hline
 Expression &~~&Description &~  &SET 1 &SET 2 \\
\hline 
\hline

$U$, meV & & Coulomb Charging energy && 4.0 & 2.5 \\
$\alpha_G$ & & Relative Capacitance of the Gate electrode && 0.065 & 0.034\\
$\Gamma$, meV & & 
\begin{minipage}[t]{ 8 cm}%
Sum of the dot-source and dot-drain single particle tunneling rates , estimated from the width of Coulomb charging peaks away from zero bias
\end{minipage} 
&& 1.6 & 0.9 \\
$T_K$, mK & & Kondo temperature at the center of the valley && 420 & 650 \\

$\frac{\sqrt{\Gamma U}}{2k_B}~ e^{(-\frac{\pi}{4}\frac{U}{\Gamma})}$, mK & &
\begin{minipage}[t]{ 8 cm}%
Mid-valley Kondo Temperature value predicted by Haldane (Eq. \ref{eq:haldane})
\end{minipage} && 1300 & 740 \\
A & & & & 0.32 & 0.88 \\
\hline
\end{tabular}
\caption[table]{\label{tab:1}Parameters of the Kondo states in SET1 and SET2}
\end{table}

\begin{figure}[H]
\includegraphics[width=5in, keepaspectratio=true]{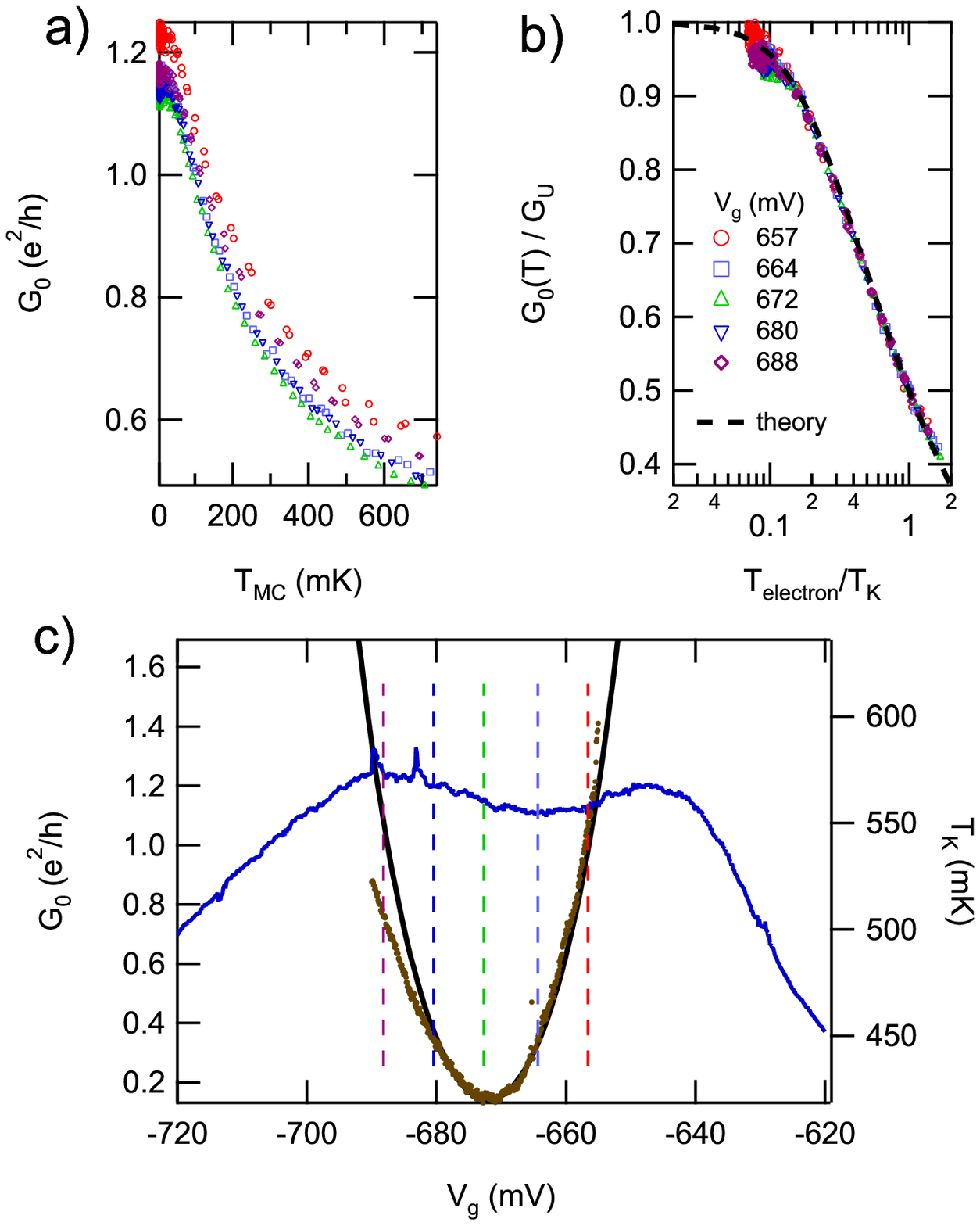}
\caption{\label{fig:1} Kondo temperature analysis for SET 1. (a) Conductance $G_0$ in SET1 as function of temperature at several representative values of $V_G$. (b) Conductance scaled by the low-temperature conductance value as function of temperature scaled by $T_K$. The black dashed line shows theoretical calculation based on eqs. 3 and 4.  (c) black trace: Conductance measured  at zero $V_{ds}$ ( left axis) as function of  $V_G$ and the Kondo temperatures obtained from fits of T-dependent conductance data ( brown symbols) and the Kondo temperature calculated  via the Haldane formula (eq. 1, black solid line, right axis) Vertical dashed lines indicate values of $V_G$ chosen for plots (a) and (b). }
\end{figure}

\begin{figure}[H]
\includegraphics[width=5in, keepaspectratio=true]{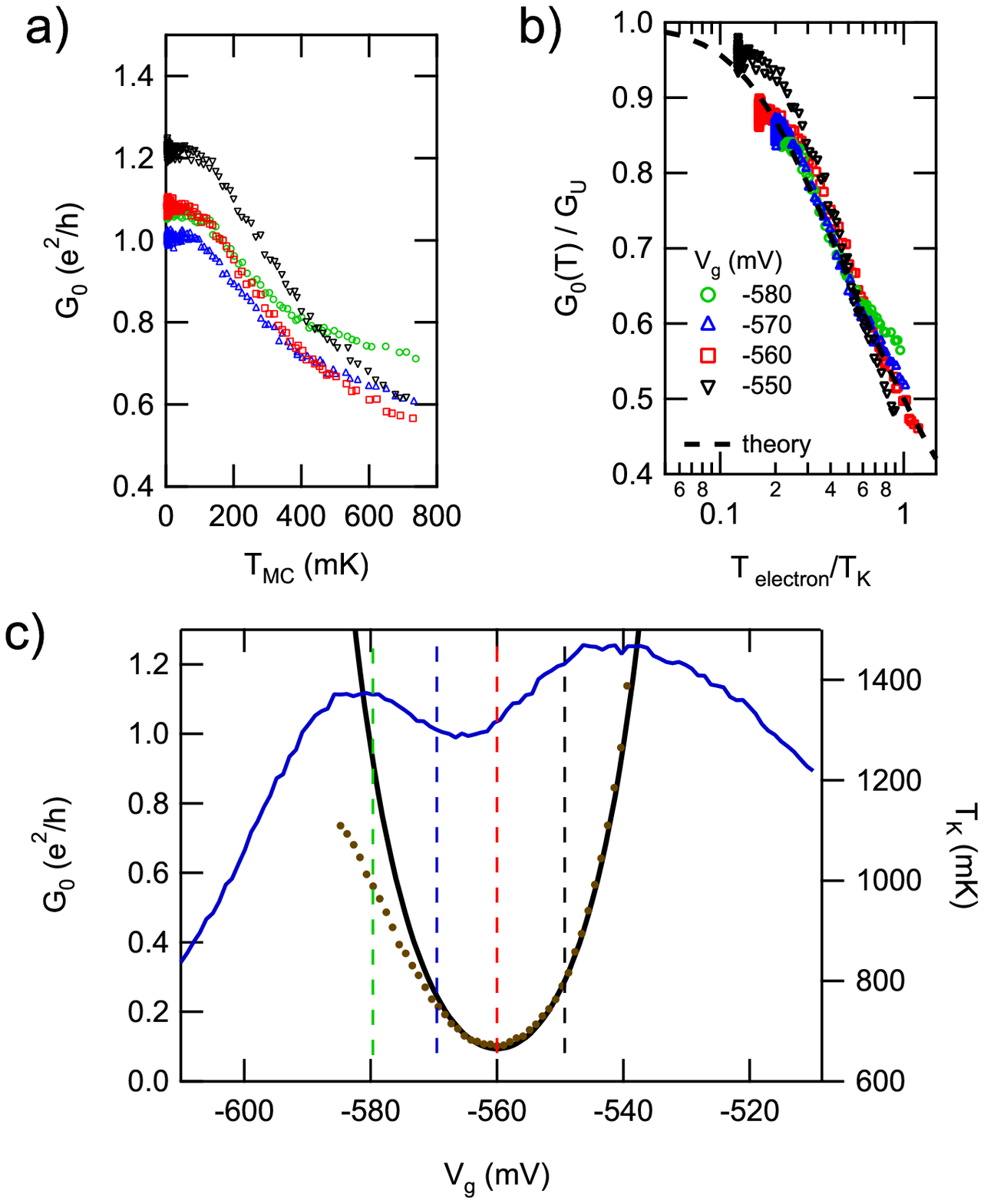}
\caption{\label{fig:2} Kondo temperature analysis for SET2. The choice of symbols and notations are identical to those in Figure 1.}
\end{figure}

\bibliographystyle{apsrev}
\bibliography{MWref_paper2}